# Honeypots for employee information security awareness and education training: A conceptual EASY training model

Lek Christopher, Kim-Kwang Raymond Choo, Ali Dehghantanha

Abstract— The increasing pervasiveness of internet-connected systems means that such systems will continue to be exploited for criminal purposes by cybercriminals (including malicious insiders such as employees and vendors). The importance of protecting corporate system and intellectual property, and the escalating complexities of the online environment underscore the need for ongoing information security awareness and education training and the promotion of a culture of security among employees. Two honeypots were deployed at a private university based in Singapore. Findings from the analysis of the honeypot data are presented in this paper. This paper then examines how analysis of honeypot data can be used in employee information security awareness and education training. Adapting the Routine Activity Theory, a criminology theory widely used in the study of cybercrime, this paper proposes a conceptual Engaging Stakeholders, Acceptable Behavior, Simple Teaching method, Yardstick (EASY) training model, and explains how the model can be used to design employee information security awareness and education training. Future research directions are also outlined in this paper.

1. **Introduction**

Information and communications technologies (ICT) form the backbone of many aspects of the critical infrastructure sectors, particularly in technologically advanced countries such as Australia and Singapore. For example, the moment a system is connected to a public facing IP address (cyberspace), the system will be probed, scanned or compromised almost immediately, as revealed in a 2014 study[1].

The challenges of securing a corporation's information security are increasingly interdisciplinary and multifaceted, as information security is defined not only by people, process and technical perfection but rather by an ability to manage these imperfections. In addition, information security threats will continue to evolve into new forms. Technical solutions, while effective, cannot provide a comprehensive solution [2]. The Australian Signals Directorate[3] noted that

*even the best technical security measures can be defeated by inappropriate user behaviour. Some users, in particular individuals and small businesses, are more vulnerable due to a general lack of awareness of cyber threats and relatively low resources devoted to information security.*

Human factors are likely to remain one of the weakest links in attempts to secure systems and networks [4]. Although information security management is relatively mature, comprehensive statistics on patterns and trends in malicious cyber activities particularly incidents involving private sector corporations, remain an elusive goal (e.g. due to unreported and undetected activities) [5].

One typical information security strategy is to create conditions unfavorable to security incidents, for example, by identifying, manipulating and controlling the situational or environmental factors to limit the opportunities for offenders to engage in criminal behavior. The routine activity theory, for example, explains that crime is generally opportunistic, which occurs when a suitable target is in the presence of a motivated offender and is without a capable guardian [6]. Therefore, to reduce security incidents, one could target each of these areas – (1 and 2) increasing the effort required to offend and the risk of getting caught, as well as (3) reducing the rewards of offending. One such strategy is employee information security awareness and education training (and user education is also one of the key cybercrime mitigation strategies identified by the Australian Signals Directorate)[7].

Employee information security awareness and education are critical in mitigating cyber threats as well as maintaining up-to-date knowledge of cybercriminal activities and mitigation measures available (e.g. to harden their systems, control access to facilities, and deflecting offenders). Major corporations generally include information security awareness program in their corporation training. Effective awareness and education training can potentially reduce the numbers of vulnerable systems that can be exploited. However, fearing negative publicity and the resulting competitive disadvantage due to leakage of information about security breaches and monitoring metrics, information about such breaches (e.g. how they occur and what are the lessons learnt) is generally not shared with most employees.

In addition, post-incident forensic investigations are not generally conducted [8]; hence, corporations may not have a complete picture of the security incident. Honeypots can be a useful tool to facilitate understanding of attack trends and attacker behaviours, as these systems are designed to emulate services of a typical server and capture attacker activities.

A number of studies [9,10,11] have highlighted the usefulness of honeypots in information security education, but they are usually discussed in the context of the higher education institution settings (i.e. educating university students) rather than in a real-world corporate environment. In this paper, we study the feasibility of deploying honeypots with the aim of collecting attacker trends, behaviours and activities that can be used in information security awareness and education training in a corporate environment.

Both honeypots in our study were deployed on a low-cost embedded devices in a Private University based in Singapore. Applying the Routine Activity Theory, we outline an employee information security awareness and education training that uses data collected from the honeypots.

The rest of the paper is organized as follows. The next section introduces our experiment setup. Findings from the Dionaea and Kippo were presented in Sections 3 and 4 respectively. We then describe how the routine activity theory can be used to design employee information security awareness and education training in Section 5. The last section concludes this paper.

2. **Experiment Setup**

We set up two honeypots, namely Dionaea [12] and Kippo [13], using Raspberry Pi (see Table 1) running on Raspbian, which is a Debian-based operating system.

Table 1  SPECIFICATIONS

| Raspberry Pi Model B | |
|---|---|
| Operating System | Raspbian |
| Processor | ARM1176JZF-S 700 MHz |
| Graphical User Interface card | Broadcom VideoCore IV GPU |
| RAM | 512MB |
| Video Out | Composite (PAL and NTSC), HDMI or Raw LCD (DSI) |
| Audio Out | 3.5mm Jack or Audio over HDMI |
| Storage | SD/MMC/SDIO |
| Networking | 10/100 Ethernet (RJ45) |
| Low-Level Peripherals: | • 8 x GPIO<br>• UART<br>• I2C bus<br>• SPI bus with two chip selects<br>• +3.3V<br>• +5V<br>• Ground |
| Power Requirements: | 5V @ 700 mA via MicroUSB or GPIO Header |

Dionaea is a low interaction honeypot which emulates services in order to collect malware targeting protocols such as Server Message Block (SMB). The SMB implementation on Dionaea is based on python and SMB emulation is based on mwcollectd. Dionaea also supports the uploading of files into the SMB shares, and protocols such as:
- Hpertext Transfer Protocol (HTTP) although it does not capture any data transmitted via HTTP,
- File Transfer Protocol (FTP) to create directories and allow the uploading and downloading of files,
- Trivial File Transfer Protocol (TFTP) for file transfer, Microsoft SQL Server (MSSQL), and
- Voice over IP (VoIP) protocol to capture incoming SIP messages.

We make use of the libemu library [14] in the capturing of malware for our study. The medium interaction Kippo honeypot collects information about attempted brute force activities targeting SSH services.

Both honeypots were setup in a network zone protected by the firewall – see Figure 1. As both systems are running low interaction and medium interaction honeypots, there is minimal risk of either honeypot being used as proxy to compromise other connected systems. The binary and logs are sent to a reporting server running on a virtual machine for further analysis. The data are extracted via SQLite for Dionaea and Mysql for kippo.

3. **Findings : Dionaea**

This honeypot was hosted from a local Internet service provider link, with a total of 115,882 IP connections to the honeypot. The data was collected over a period of 97 days, from 25 Oct 2013 to 30 Jan 2014.

Figure 2 shows the breakdown of targeted ports on the Dionaea honeypot. The highest number of hits (77.52%) was on port 445 (SMB, the main protocol run by Dionaea). This is followed by port 139 (15.21%) and port 135 (5.21%).

*3.1 Breakdown of Attacks by Time of Day*

Figure 3 shows the distribution of connection over the 24-hour period. It was noted that connections followed a 9 am to 5 pm GMT+8, peaking at 12 pm and a sharp drip at 5 pm.

This could be because more machines are connected online during a typical work day cycle.

*3.2 Breakdown of Attacks by IP Address*

Table 2 presents a breakdown of detected successful connections by IP addresses. Of the six countries associated with the top ten IP addresses, United Kingdom (UK) accounted for close to three-quarter of detected successful connections. It was noted that this result does not imply that there are more attackers from UK, as the IP address may be spoofed or belonged to a compromised computer based in the country. The spread of the different geo-location spread does, however, suggest that attacks could have originated from any part of the world.

Table 2  BREAKDOWN OF DETECTED SUCCESSFUL CONNECTIONS BY COUNTRIES (DIONAEA HONEYPOT)

| Connections | IP | Country |
|---|---|---|
| 58609 | 176.227.xxx.xxx | United Kingdom, England, Gosport |
| 4957 | 2.95.xxx.xxx | Russian Federation, Moscow City, Moscow |
| 4937 | 78.61.xxx.xxx | Lithuania, Vilniaus Apskritis, Vilnius |
| 2972 | 186.89.xxx.xxx | Venezuela, Bolivarian Republic of, Miranda, Petare |
| 2627 | 186.89.xxx.xxx | Venezuela, Bolivarian Republic of, Miranda, Petare |
| 2253 | 221.143.xxx.xxx | Korea, Republic of, Seoul-t'ukpyolsi, Seoul |
| 1624 | 2.95.xxx.xxx | Russian Federation, Moscow City, Moscow |
| 1334 | 93.183.xxx.xxx | Bulgaria, Khaskovo, Khaskovo |
| 837 | 2.95.xxx.xxx | Russian Federation, Moscow City, Moscow |
| 753 | 212.21.xxx.xxx | Bulgaria, Grad Sofiya, Sofia |

Further analysis was carried out on the actual number of malware downloaded from the connections. As noted in Table III, although the top connection was from Gosport in United Kingdom (a total of 58,609 connections), there were only 962 instances of malware download. The high number of connections could be for reconnaissance purpose prior to the delivery of the malicious payload. It was also noted that connections originating from Korea did not result in any malware download. Similar, these connections could be probing for specific information about the host prior to proceeding with an actual exploitation (which may not involve the use of malware).

Table 3 FURTHER ANALYSIS OF DOWNLOADS FROM TOP 10 CONNECTIONS BY COUNTRIES

| Downloads | IP | Country |
|---|---|---|
| 962 | 176.227.xxx.xxx | United Kingdom, England, Gosport |
| 2400 | 2.95.xxx.xxx | Russian Federation, Moscow City, Moscow |
| 2425 | 78.61.xxx.xxx | Lithuania, Vilniaus Apskritis, Vilnius |
| 1423 | 186.89.xxx.xxx | Venezuela, Bolivarian Republic of, Miranda, Petare |
| 1271 | 186.89.xxx.xxx | Venezuela, Bolivarian Republic of, Miranda, Petare |
| 0 | 221.143.xxx.xxx | Korea, Republic of, Seoul-t'ukpyolsi, Seoul |
| 794 | 2.95.xxx.xxx | Russian Federation, Moscow City, Moscow |
| 652 | 93.183.xxx.xxx | Bulgaria, Khaskovo, Khaskovo |
| 413 | 2.95.xxx.xxx | Russian Federation, Moscow City, Moscow |
| 358 | 212.21.xxx.xxx | Bulgaria, Grad Sofiya, Sofia |

*3.3 Malware Captured by Dionaea*

The majority of the captured malware from this honeypot were Conficker worm[15] exploiting the RPC vulnerability MS08-067 [16]. Figure 4 shows the breakdown of the different variant of Conficker worm captured. The presence of Conficker proved its resiliency since it was first discovered in 2008. For example, a report by F-Secure noted that one third of detected threats were attributed to Conficker[17].

4. **Findings : Kippo**

The dataset was collected over a period of 483 days, between 18 Aug 2013 and 14 Dec 2014. A number of **1,075,161** login attempts to the Kippo honeypot was recorded, and of which, **3,762** single unique IP addresses were logged.

*4.1 Top Ten Passwords Attempted*

Passwords are analogous to our key to a locked door, and weak usernames and passwords are often a vector that can be targeted by attackers to gain entry into the system.

The most commonly passwords attempted were "admin", and the different variants of "password" – see Figure 5. Other popular attempted passwords include common words (e.g. apple - 1706, pass - 1565, user - 1017), common names (e.g. David - 551, Peter - 316, Sally - 202), and common keyboard stroke (e.g. qwerty - 2110, qazwsx - 2111, q!w@e#r$ - 71).

The length of the passwords attempted generally ranged from six to nine characters (i.e. >50% of attempted passwords).

The two most commonly used username are root and admin. The former is the default username for Unix based machine while the latter is a commonly used username in routers or network devices. It was noted that oracle was among the top 10 username as it is a widely used database. Our findings echoed studies such as [21,22].

Most organizations will not have a password policy that requires a long password [23], and various studies (see [24]) recent high profile breaches suggested that reusing usernames and passwords is not an uncommon practice among users. For example, a recent study found several hundred thousand leaked passwords from eleven web sites that "43-51% of users reuse the same password across multiple sites" [25]. Based on the findings, the authors designed a 'cross-site password-guessing algorithm, which is able to guess 30% of transformed passwords within 100 attempts compared to just 14% for a standard password-guessing algorithm without cross-site password knowledge'. Another recent study by Lu, Shuai and Yu [26] also demonstrated that it is relatively easy to identify online individual e-commerce customers by using the customers' username.

*4.2 Top Ten IP Connections*

The chart below outlined the top ten unique IP addresses visiting the honeypot, where majority of the IP connections were from Hong Kong and China. This is a similar observation reported by Cisco [18] but differs from recent studies undertaken by researchers from Aristotle University of Thessaloniki[19] and University of Ostrava [20]. In the first research, a Dionaea honeypot was deployed between 19 February and 11 March 2012, 28 March and 23 April 2012, and 21 January and 19 February 2013. Similar to our research, Dionaea and Kippo honeypots were deployed. The top five attack countries varied between the three deployment periods in the first research. Unlike our findings, the top attack countries in both studies are dominated by European countries. This is, perhaps, due to the location of the deployed honeypots.

**4.3 Top Ten Successful Commands**

Kippo simulates a system with limited command line. It allows the pre-configuration of username and password for login. In this setup, we used the default username "root" and password "admin" as the login username and password in order to gain a better understanding of attacker activities once they are logged into the compromised system. Such activities are captured in the database and allow real-time playback in as shown in Figure 8.

Figure 9 listed the top ten commands executed when an attacker was connected to the system. It was observed that the folder of interest appears to be /tmp, which stores temporary files that will automatically be deleted upon boot up. Attackers were observed to upload their artifacts or tools to the /tmp folder. The second most popular activity is the directory listing command, ls, which provides the attacker an overview of the system file structure.

*4.4 Files downloaded*

There were a total of 93 unique files downloaded as shown in Figure 10. We observed that a majority of the binaries are designed for 32-bit executables on MIPS or Intel platform. A handful of the captured binaries were compiled for ARM chipset.

Further analysis of the captured malware shows that a significant number of the binaries were designed to conduct denial of service attacks. The other popular captured malware were backdoors targeting Linux platform, which reflects the trend highlighted in the 2014 McAfeee Labs Threat Report [27].

5. **A conceptual EASY training model**

Most major corporations would have existing information security awareness and education training in place. The design of such training could be customized for individual corporate, although it would generally consist of three basic building blocks, namely: Awareness, Training and Education [28].

- *Awareness typically begins with ensuring all employees within the organization with the basic cybercrime understanding and the importance of information security, generally achieved via ongoing training and education – the other two building blocks.*
- *During training, participants (e.g. employees) will be taught relevant and up-to-date skills and competencies in order to contribute to a culture of security within the corporate.*
- *As aptly summarized in [29], education integrates all essential skills and relevant competencies into a common body of knowledge for information security specialists and professionals.*

Figure 11 describes the relationship between the three building blocks.

Enhancing or improving information security management in corporate has been studied by researchers. For example, Nersen, Rana, Mumtaz [30] identified several factors that could be included in awareness and education training. In another recent work, Martini and Choo [31] explained how the Situational Crime Prevention, a criminology theory, can be used as the underlying theoretical lens in the design of cybersecurity courses.

The Routine Activity Theory [32] is another popular criminology theory that has been widely used to study cybercrime. The theory states that for crime to occur, there must be three elements, namely a person motivated to commit the offense, a vulnerable victim who is available, and insufficient protection to prevent the crime. The theory draws on rational exploitation of 'opportunity' in the context of the regularity of human conduct to design crime prevention strategies, especially where terrestrial interventions are possible. Therefore, to reduce the probability of the occurrence of a crime, we would need to

- Increase the effort required to offend;
- Increase the risk of getting caught; and/or
- Reduce the rewards of offending.

Our proposed **E**ngaging Stakeholders, **A**cceptable Behavior, **S**imple Teaching method, **Y**ardstick (EASY) training model can be used with the Routine Activity Theory to design training activities, which will increase the effectiveness of security awareness training and enhance the security culture within the corporate – see Table 4.

Table 4    APPLYING ROUTINE ACTIVITY THEORY TO THE CONCEPTUAL EASY MODEL

| EASY Model | Routine Activity Theory | | |
| --- | --- | --- | --- |
| | Increasing the effort required to offend | Increasing the risk of getting caught | Reducing the rewards of offending |
| **Engaging Stakeholders** | | | |
| Engage senior management support | Yes | Yes | |
| Engage employee (including contractors and vendors) | Yes | Yes | |
| Engage with industrial partner s | Yes | Yes | |
| **Acceptable Behavior** | | | |
| Reward Good User Behavior | Yes | | Yes (particularly for insider-related threats) |
| Improved social support | | Yes | |
| Improved feedback and assistance | | Yes | |
| User ownership in detection and suspected incidents, compromises or anomalies | Yes | Yes | |
| **Simple Teaching Method** | | | |
| Explore effective training methods | Yes | | |
| Scalable and cost effective | | Yes | |
| **Yardstick** | | | |
| Measurement through assessment | Yes | | |
| Multiple channel feedback | Yes | | |

We now explain the four pillars of the model, as well as how honeypot data can be used in the model.

*5.1 Engaging Stakeholders*

A successful security awareness program requires commitment and support of all stakeholders, including senior management (e.g. the C-level executives). For example, a committed supportive senior management team who is able to lead by example (e.g. active participants in corporate training activities) will reinforce the importance of information security and set the right tone in ensuring a healthy security culture within the corporate. In addition, a clearly committed senior management will help ensure that the corporate's information security training program is adequately funded and implemented.

Employees are generally the front-line of information security threats and, therefore, active participation from employees will play a key role in early detecting and mitigation of information security threats (e.g. reporting of abnormalities which will lead to follow-up investigation and formulation of mitigation strategies). This would significantly increase the effort for an attacker to gain entry, or increase the risk of detection. Security incident response teams should also consider join global alliances, such as Anti-Phishing Working Group (APWG) and Forum for Incident Response and Security Teams (FIRST), to facilitate timely sharing of information to better combat security threat.

Traditionally, security awareness training materials are delivered using presentation slides and e-learning platform, which are generally non-engaging and information may not be up-to-date. The use of honeypot data could contribute to this gap.

In our model, we propose using interactive tools such as data visualization tools, as the latter has been shown to be an effective way to positively influence user [33]. Using real data collected from honeypots, materials delivered using data visualization tools can help ensure employees have an up-to-date and in-depth understanding of attacker trends. For example, the top attacks ports collected from both honeypots would allow participants to understand:

- Attack vectors targeted by attackers (e.g. see Figures 2 to 4, and 10);
- Attack origins and time of attacks (e.g. see Figure 3 and Tables 2 and 3);
- Commonly attempted username and password combinations (e.g. see Figures 5 to 7); and
- Attacker activities on the system (e.g. see Figures 8 and 9).

Such information would also allow the corporate to identify their vulnerable systems that can be exploited and ensure appropriate security measures are in place (e.g. turning off services that are not required to reduce the exposure). Employees would also be able to understand that attacks do not originate from one country or region, and attacks can originate from countries as far as Lithuania (in the context of the honeypot setup in this study). In addition, honeypot data would provide employees with up-to-date information on malware trends (e.g. what are the current malware targeting the corporate).

Honeypot data could also be used to provide senior management with an overview of the threat landscape (e.g. via a honeymap) as well as a real-time data visualization attack traffic to demonstrate how corporate systems are being targeted by attackers located in different places around the world (see Figure 12).

*5.2 Acceptable Behavior*

An important element to ensure the success of a security awareness initiative is to inculcate the correct user behavior towards information security. Every employee has their specific role and responsibilities within the company. We need to build a culture of ownership towards the responsibility of security among all employees. This could start with equipping employee with a good understanding of current cybercrime threats and trends (e.g. information about the latest phishing scam which may results in employee system being compromised or their account credentials phished), the potential impact of a security breaches, and mitigation strategies.

Employees should also be educated about the importance of taking personal responsibility and ownership to ensure a secure environment and adopt cyber hygiene practices (e.g. not sharing username and password). This is also consistent with the Australian Government information security manual, which emphasized the importance of ensuring employees are familiar with their roles and responsibilities, understand and support security requirements, and learn how to fulfill their security responsibilities [35].

Employee could be motivated using an appropriate reward system for employees demonstrating competency and walking the talk on security best practices. For example, we could organize campaign to publicly identify the 'most information security aware employee of the month' who demonstrates such values through nomination. We could also incorporate the use of honeypot to influence user behavior. For example, we could demonstrate to end user how rapidly malware could infect Dionaea honeypot when it is connected to the internet and provide preview through the playback action of what was done by attacker upon intrusion. We attempt to immerse user through live "hack" scenario and this could enrich user learning experience which was not found in conventional awareness training programme.

Insider abuse has been the subject of academic research in the last few decades, but remains a concern area to corporates and governments. For example, Verizon data breach report 2014 [34] identified that the third most frequent data breach was caused by insider misuse. One deterrence strategy against insider abuse is to publish anonymized audit findings, for example, findings of unauthorized access by employees, to raise awareness and deter future offending. In addition, corporates should consider introducing employee counseling initiatives that include avoiding having disgruntled employee carrying out malicious act. This is enhanced by providing appropriate social support within employee to establish a positive environment condoning bad behavior and allow appropriate feedback channel for whistle blowing.

Security is often being regarded as the responsibility of the information technology and information security division within the corporation. Given the limited resources and wide threat surface, there are multiple entry points to a corporate system that can be exploited by cybercriminals. Therefore, it is important to put in place reporting requirements for employees to report detected and suspected incidents, compromises or anomalies. In addition, having an established corporate communication channel (e.g. an online incident reporting platform) and documentation of security best practices will enhance employee information security awareness and knowledge, and this could be reinforced using mock exercises with up-to-date data collected from honeypots.

*5.3 Simple Teaching method*

Senior stakeholders in the organizations need to understand the importance of the right governance enablers and more importantly, to understand that information security is not only a cost or an IT issue, but it can facilitate economic exchange and deliver real business benefits. It is important for senior stakeholders in the organizations to be able to answer questions such as

1. Who would benefit from having access to our information and systems?
2. What makes us secure against security threats?
3. Is the behaviour of my staff enabling a strong security culture?
4. Are we ready to respond to an information security incident?
5. What would a serious information security incident cost our corporate?
6. How much effort will be needed to mitigate and recover from a serious cyber security incident?

The teaching method of the security awareness training must be simple and easy to use. Common methods include classroom-based, computer-based e-learning, and video-based learning, each requires different considerations on cost and scalability – see Figure 13.

It is also recommended that corporates consider putting classroom-based security training for new employees, and re-enforcement training be delivered via video and/or e-Learning based. Regardless of the delivery method, the principle of the teaching method should be simple, direct to the point and easily understood by user with different background. Avoid using technical jargons and terminologies, which may not be easily understood by the employee. As previously explained, the training materials could be reinforced using live examples (e.g. live captured activities from the honeypot) to walk participants through the different aspect of security best practices.

We also design a simple password checking application (see Figure 14) where participants could key in their password to check against the data collected from the honeypots and other known sources (e.g. known databases of compromised username and passwords) whether their password is one of those commonly attempted by attackers. This is a simple and cost-effective way to self-assess password as well as reinforcing the importance of having a strong password.

*5. 4 Yardstick*

For any program to be successful, we need to be able to measure its outcome and monitor the progress. The monitoring should be done continuously so that gaps could be identified early and appropriate remediation could be applied. Therefore, an appropriate feedback channel should be established such that improvement to the security awareness training could be made. This could be provided in the form of questionnaire surveys, evaluation forms, interviews and focus group, and success criteria need to be established (e.g. employee attendance rate and minimal score to pass assessment test).

**6. Conclusion and future work**

Honeypot data can provide useful information which could be used in the understanding of attacker trends during employee information security awareness and education training. A better insight and knowledge of attacker trends will guide further responses at the operational level (e.g. effectiveness of existing controls), and contribute to management policy making and reform within the corporate.

- At a strategic level, will inform and help senior management and other key stakeholders to reach a level of consensus and decide on broad strategies, policies and resources for the corporate in a timely fashion.
- At the operational level, findings about existing and emerging patterns of network activities will support policy makers and other key stakeholders in their decisions about focusing scarce resources in the most effective way (e.g. effect change to harden the environment).

It is also likely that corporations would be more likely to share such data internally considering the non-sensitive nature of the data, thus, allowing all employees to have up-to-date intelligence and ensuring that appropriate controls.

In this paper, we illustrated how the Routine Activity Theory can be used as the underlying theoretical lens to design an employee information security awareness and education training, which incorporates real-world attack data (collected from honeypots). This approach has the potential to raise information security awareness within an organization as well as establishing a culture of security within the organization, and consequently, increase the effort required to offend, increase the risk of getting caught and reduce the rewards of offending.

Future work would include deploying honeypots in corporations and institutions in different countries located at different regions, as well as customized honeypot tools for specific security awareness themes (e.g. mobile security, web applications security, and industrial control system). We will also collaborate with like-minded researchers and practitioners to deploy our conceptual EASY model. This would allow us to receive practical feedback on the suitability of the various pillars in the model, etc, which will provide the basis for the best practice (and the library of training methodologies and tools) recommendations. Face-to-face interviews will also be conducted with relevant stakeholders to determine the feasibility of the refined model and library of training methodologies and tools.

*Competing Interests*

The authors declare that they have no competing financial or commercial interests.

*Authors' Contributions*

CSCK and KKRC designed the experiment setup. CSCK conducted the technical experiments and collected the findings. CSCK and KKRC analyzed the findings and designed the conceptual EASY training model. All three authors (CSCK, AD and KKRC) drafted the manuscript, read and approved the final manuscript.

*Acknowledgment*

The views and opinions expressed in this article are those of the authors alone and not the organizations with whom the authors are or have been associated / supported. The authors would also like to thank the Private University in Singapore for deploying the two honeypots used in this study.